\documentclass[11pt]
{amsart}
\usepackage{pdfpages}
\usepackage{hyperref}
\hypersetup{nesting=true,debug=true,naturalnames=true}
\usepackage{graphicx,amssymb,upref}

\newtheorem{theorem}{Theorem}
\newtheorem{lemma}[theorem]{Lemma}
\newtheorem{definition}{Definition}

\title{Ebert's Asymmetric Three Person Three Color Hat Game}
\author{Theo van Uem}  
\address{Amsterdam University of Applied Sciences, Amsterdam, The Netherlands.} 
\email{tjvanuem@gmail.com}  
\begin{document}
\hbadness=99999

\begin{abstract}
 We generalize Ebert’s Hat Problem for three persons and three colors. All players guess simultaneously the color of their own hat observing only the hat colors of the other players. It is also allowed for each player to pass: no color is guessed. The team wins if at least one player guesses his or her hat color correct and none of the players has an incorrect guess. This paper studies Ebert’s hat problem, where the probabilities of the colors may be different (asymmetric case). Our goal is to maximize the probability of winning the game and to describe winning strategies. In this paper we use the notion of an adequate set. The construction of adequate sets is independent of underlying probabilities and we can use this fact in the analysis of the asymmetric case. Another point of interest is the fact that computational complexity using adequate sets is much less than using standard methods.
\end{abstract}
\maketitle
\section{Introduction}
Hat puzzles were formulated at least since Martin Gardner’s 1961 article~\cite{MG}. They have got an impulse by Todd Ebert in his Ph.D. thesis in 1998~\cite{TE}. Buhler~\cite{JB} stated: “It is remarkable that a purely recreational problem comes so close to the research frontier”. Also articles in The New York Times~\cite{SR}, Die Zeit~\cite{WB} and abcNews~\cite{JP} about this subject got broad attention. This paper studies a generalization of Ebert’s  hat problem. The original problem was symmetric: each color has probability $\frac{1}{3}$. We consider the asymmetric case: three distinguishable players are randomly fitted with a colored hat (three colors available), where the probabilities of getting a specific color  may be different, but known to all the players. All players guess simultaneously the color of their own hat observing only the hat colors of the other two players. It is also allowed for each player to pass: no color is guessed. The team wins if at least one player guesses his or her hat color correctly and none of the players has an incorrect guess. Our goal is to maximize the probability of winning the game and to describe winning strategies.
The symmetric two color hat problem (probability $\frac{1}{2}$ for each color) with $N=2^k-1$ players is solved in~\cite{EMV}, using Hamming codes, and with $N=2^k$ players in~\cite{GC} using extended Hamming codes. 
Burke et al.~\cite{EB} try to solve the symmetric hat problem with $N=3,4,5,7$ players using genetic programming. Their conclusion: The $N$-prisoners puzzle (alternative names: Hat Problem, Hat Game) gives evolutionary computation and genetic programming a new challenge to overcome. 
Lenstra and Seroussi~\cite{HL} show that in the symmetric case of two hat colors, and for any value of $N$, playing strategies are equivalent to binary covering codes of radius one.
Combining the result of Lenstra and Seroussi with Tables for Bounds on Covering Codes~\cite{GK}, we get:\par
\begin{center}
\begin{tabular}{|c|c|c|c|c|c|c|c|c|}
\hline
$N$&2&3&4&5&6&7&8&9\\
\hline
$K(N,1)$&2&2&4&7&12&16&32&62\\
\hline
\end{tabular}
\par
\end{center}

$K(N,1)$ is smallest size of a binary covering code of radius 1. Maximum probability for Ebert's symmetric two color Hat Game is $1-\frac{K(N,1)}{2^N}$. Lower bound on $K(9,1)$ was found in 2001 by Östergård-Blass, the upper bound in 2005 by Östergård.
Krzywkowski~\cite{MK} describes applications of the hat problem and its variations, and their connections to different areas of science. Krzywkowski~\cite{MKR} and Guo et al. ~\cite{GU} gives an optimal solution of the symmetric three person three color hat  problem. Tantipongpipat~\cite{TA} proves an optimal strategy for Ebert’s hat game with three players and more than two hat colors.
Johnson~\cite{BJ}  ends his presentation with an open problem:
If the hat colors are not equally likely, how will the optimal strategy be affected?
We will answer this question and our method gives also interesting results in the symmetric case.
In section 2 we define an adequate set. 
In section 3 we obtain results for the asymmetric three person three color Hat Game, where each player has the same probabilities to get a specific colored hat.
In section 4 we get new results for the symmetric three person three color Hat Game by using the adequate set method.
In all situations all players know the underlying probabilities of each player.

\section{Adequate sets}
In this section we have $N$ players and $q$ colors.
The $N$ persons in our game are distinguishable, so we can label them from 1 to
$N$. We label the $q$ colors $0,1,..,q-1.$ The probabilities of the colors are
fixed and known to all players. The probability that color $i$ will be on a hat is
$p_i$ $(i\in\{0,1,..,q-1\},\ \ \sum_{i=0}^{q-1}p_i=1).$
Each possible configuration of the hats can be represented by an element of
$B=\{b_1b_2\dots b_N\vert b_i\in\left\{0,1,\dots,q-1\right\},\ i=1,2..,N\}$.
The S-code  represents what the $N$ different players sees. Player $i$ sees q-ary code $b_1..b_{i-1}b_{i+1}..b_N$ with decimal value
$s_i=\sum_{k=1}^{i-1}b_k.q^{N-k-1}+\sum_{k=i+1}^Nb_k.q^{N-k}\ $, a value between 0 and $q^{N-1}-1.$ \\
Let S be the set of all S-codes:
$S=\{s_1s_2\dots s_N\vert{}s_i=\sum_{k=1}^{i-1}b_k.q^{N-k-1}+\sum_{k=i+1}^Nb_k.q^{N-k},b_i\in{}\{0,1,\dots,q-1\},\
i=1,2,\dots,N\ \}$.
Each player has to make a choice out of $q+1$ possibilities: 0='guess color 0', 
1='guess color 1', \ldots{}.,$\ q-1$ ='guess color $q-1$', $q$='pass'.
\\
We define a decision matrix $D=\left(a_{i,j}\right) \ $ where
$i\in{}\{1,2,..,N\}$(players); $j\in{}\{0,1,..,q^{N-1}-1\}$(S-code of a player);
$a_{i,j}\in{}\left\{0,1,..,q\right\}.$
The meaning of $a_{i,j}$ is: player $i$ sees S-code  $j$ and takes decision $a_{i,j}$ (guess a color or pass).
We observe the total probability (sum) of our guesses.
 
For each $b_1b_2\dots b_N$ in B  with $n_i$ times color $i$   $(i=0,1,\dots,q-1,
\sum_{i=0}^{q-1}n_i=N$) and S-code   player $i$:
$s_i=\sum_{k=1}^{i-1}b_k.q^{N-k-1}+\sum_{k=i+1}^Nb_k.q^{N-k}$ we have:\newline

CASE  $b_1b_2\dots b_N$  \\
IF $a_{1{,s}_1}\in{}\{q,b_1\}$  AND $a_{2,s_2}\in{}\{q,b_2\}$  AND ... AND 
$a_{N,s_N}\in{}\{q,b_N\}$  AND \\ NOT
($a_{1,s_1}=a_{2{,s}_2}=\dots=a_{N,s_N}=q)$  THEN sum=sum+$p_0^{n_0}.p_1^{n_1}\dots\
p_{q-1}^{n_{q-1}}$.
\newline

Any choice of the $a_{i,j}$ in the decision matrix determines which CASES $b_1b_2\dots b_N$ have a
positive contribution to sum (we call it a GOOD CASE) and which CASES don't
contribute positive to sum (we call it a BAD CASE).
\begin{definition}
 Let $A \subset B$. $A$ is adequate to $B-A$ if for each q-ary element $x$ in $B-A$ there are $q-1$ elements in A which are equal to $x$ up to one fixed q-ary position.
 \end{definition}
 
\begin{theorem}
 BAD CASES are adequate to GOOD CASES.
\end{theorem}
\begin{proof}
  Any  GOOD CASE  has at least one $a_{i,j}$ not equal to $q$. Let this
specific $a_{i,j}$ have value $b_{i_0}.$ Then our GOOD CASE generates $q-1\ $BAD CASES
by only changing  the value $b_{i_0}$ in any value of $0,1,..,q-1\ $
except $b_{i_0}$.  
\end{proof}
The definition of adequate set is the same idea as the concept of 
strong covering, introduced by Lenstra and Seroussi~\cite{HL}. The number of elements in an adequate set will be written as \textit{das} (dimension of adequate set).
Adequate sets are generated by an adequate set generator (ASG). For an implementation of an ASG in VBA/Excel  see Appendix \ref{appendix:one}. 
\begin{theorem}
Each adequate set generates a  decision matrix $D$.
\end{theorem}
\begin{proof}
For each element in the adequate set:
\begin{itemize}
\item Determine the q-ary representation $b_1b_2\dots b_N$
 \item Calculate S-codes   $s_i=\sum_{k=1}^{i-1}b_k.q^{N-k-1}+\sum_{k=i+1}^Nb_k.q^{N-k}$( $i$=1,..,N)
\item For each player $i$: fill decision matrix with  $a_{{i,s}_i}=b_i$  ($i$=1,..,N), where a cell may contain several values.
\end{itemize}
Matrix $D$ is filled with BAD COLORS. We can extract the GOOD COLORS by considering all $a_{i,j}$ with $q-1$ BAD COLORS and then choose the only missing
color. In all situations with less than $q-1$ BAD COLORS we pass.
When there is an $a_{i,j}$ with $q$ BAD COLORS all colors are bad, so the first
option is to pass. But when we  choose any color, we get a situation with
$q-1$ BAD COLORS.  So in case of $q$ BAD COLORS we are free to choose any color or
pass.
The code for pass is $q$, but in our decision matrices we prefer a blank, which
supports readability.
The code for `any color or pass will do' is defined  $q+1$, but in our decision
matrices we prefer a $\star$. 
\end{proof}
We call the procedure just described DMG (Decision Matrix Generator). 
An implementation of a DMG in VBA/Excel can be found in Appendix \ref{appendix:two}.

\section{Asymmetric Hat Game with three players and three colors}
In this section we obtain results for asymmetric Hat Game with three players and three
colors with  hat probabilities $p,q,r$ $(p+q+r=1, pqr>0).$
Without loss of generality we suppose $p \geq q \geq r$.
\subsection{Optimal winning probabilities}
We first execute the program ASG with values  \textit{p=0.7, q=0.2, r=0.1, \textit{das}=2,3,..,11}, which yields no adequate sets. Setting  \textit{das=12} gives a collection of 324 adequate sets.

{\raggedright
 There are 3 optimal adequate
sets, each with probability 0.242:
}

\begin{center}
  
    \begin{tabular}{rrrrrrrrrrrrr}
    4     & 5     & 7     & 8     & 9     & 13    & 14    & 16    & 17    & 18    & 20    & 24    & 0.242 \\
    1     & 2     & 8     & 12    & 13    & 15    & 16    & 20    & 21    & 22    & 24    & 25    & 0.242 \\
    3     & 6     & 8     & 10    & 11    & 13    & 14    & 19    & 20    & 22    & 23    & 24    & 0.242 \\
    \end{tabular}%
 \end{center}
 {\raggedright
So maximal winning probability of this game is 1-0.242=0.758
}

{\raggedright
The probability of an adequate set  is a  function  $\varphi=\sum{p^aq^br^{3-a-b}}$, where the
summation is over all elements of the adequate set and $a$ and $b$  are the number of zero's and one's in the ternary representation of each element of the adequate set.
}

{\raggedright
For each of the three adequate sets just found,we have:
}

{\raggedright

\[
{\varphi{}}=qr^2+2q^2r+q^3+3pr^2+2pqr+pq^2+p^2r+p^2q
\]
(e.g. the first element of the first adequate set, 4, is ternary 011 and contributes $pq^2$ to $\varphi{}$; this process is implemented in matrix  $a(i,j)$ in ASG, see appendix \ref{appendix:one}).

}

{\raggedright
We describe this as follows:
}
\begin{center}
  \begin{tabular}{|cccccccccc|}
  \hline
    00    & 01    & 02    & 03    & 10    & 11    & 12    & 20    & 21    & 30 \\
    0    & 1    & 2   & 1    & 3    & 2    & 1    & 1    & 1    & 0 \\
    \hline
    
    \end{tabular}%
\end{center}

{\raggedright
In the first line $ab$ means: we have  $p^aq^br^{3-a-b}$ and in the second line
we find the coefficients of $p^aq^br^{3-a-b}$. We call (0121321110) the pattern of the
adequate set.
\begin{definition}
 A pattern P1  is \textit{dominant} over  pattern P2 when the
$\varphi{}$ value of P1 is equal or less than the $\varphi{}$   value of P2. 
\end{definition}
%{\raggedright
The number of adequate sets when $\textit{das} \geq 12$ is shown in the following
table (use ASG with different values of \textit{das}).
%}
\begin{center}
\resizebox{\columnwidth}{!}{%
    \begin{tabular}{|c|c|c|c|c|c|c|c|c|c|c|c|c|c|c|c|c|}
    \hline
    12  & 13  & 14  & 15  & 16  & 17  & 18  & 19  & 20  & 21  & 22  & 23  & 24  & 25  & 26  & 27 \\
    \hline
   
   324     & 5832      & 50814      & 237816     & 670464      & 1194345     & 1389018     & 1097388      &  613251     & 250272     &   76086    & 17334      &   2925    &  351     & 27     & 1 \\
    \hline
  
\end{tabular}%
}
\end{center}

{\raggedright
We have to automate the search process of dominant patterns.
}
\begin{lemma}

The b-pattern
$b_1b_2...b_{10}$, where
$
b_1=b(0, 0)\  b_2=b(0, 1) \ b_3=b(0, 2) \ b_4=b(0, 3) \ b_5=b(1, 0)\  b_6=b(1, 1)\  b_7=b(1, 2) \ b_8=b(2, 0) \ b_9=b(2, 1) \ b_{10}=b(3, 0)
$
with dim(b)=$ \Sigma b_i $
{\raggedright
is dominant over the a-pattern
}
$a_1a_2...a_{10}
$ , where
$a_1=a(0, 0)\  a_2=a(0, 1)  \ a_3=a(0, 2) \ a_4=a(0, 3) \ a_5=a(1, 0)\  a_6=a(1, 1) \ a_7=a(1, 2)  \ a_8=a(2, 0) \  a_9=a(2,1)\   a_{10}=a(3, 0)
$
with dim(a)=$ \Sigma a_i $,
$dim(a)=dim(b)$,\\ 
$c_j=b_j-a_j (j=1,2,..,10), m_j=max(0,-c_j)$
 when:\\
$c_1 \geq 0 $\newline
$c_1+c_2  \geq m_3+m_5$ \newline
$(c_1+c_2+c_3  \geq m_4+m_5+m_6) \lor (c_1+c_2+c_3  \geq m_4  \land   c_5 \geq 0 \land c_5+c_6 \geq 0 )$ \newline
$(c_1+c_2+c_3+c_4  \geq m_5+m_6) \lor (c_1+c_2+c_3+c_4  \geq 0  \land   c_5 \geq 0 \land c_5+c_6 \geq 0 )$ \newline
$c_1+c_2+c_5  \geq m_3$ \newline
$(c_1+c_2+c_3+c_5+c_6  \geq m_4+m_7+m_8) \lor (c_1+c_2+c_3+c_5+c_6  \geq m_8  \land   c_4 \geq 0 \land c_4+c_7 \geq 0 )$ \newline
$c_1+c_2+c_3+c_4+c_5+c_6+c_7  \geq m_8 $ \newline
$(c_1+c_2+c_3+c_5+c_6+c_8  \geq m_4+m_7) \lor (c_1+c_2+c_3+c_5+c_6+c_8  \geq 0 \land   c_4 \geq 0 \land c_4+c_7 \geq 0 )$ \newline
$c_1+c_2+c_3+c_4+c_5+c_6+c_7+c8+c9  \geq 0 $ \newline

\end{lemma}
\begin{proof}
{\raggedright
The dominance relations between the atoms of $\varphi{}$ are (dominance is represented by an arrow):
}
\begin{center}
\includegraphics[scale=0.8]{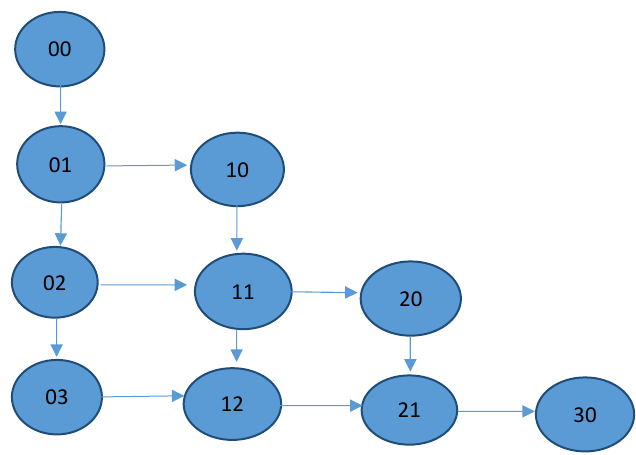} 
\end{center}

{\raggedright
The b-pattern is dominant over the a-pattern when in \emph {each} vertex of the next directed graph there is enough compensation:
}
\begin{center}
\includegraphics[scale=0.8]{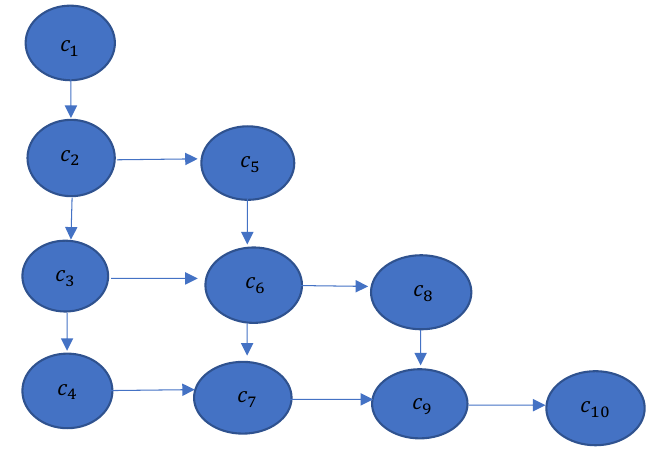} 
\end{center}
To illustrate our proof we concentrate on cell  6 (with $c_6$). Maximum flow to cell 6 (inclusive cell 6) is $c_1+c_2+c_3+c_5+c_6$. But eventually cell 3 has  to compensate a possible negative value $c_4$ in cell 4. Also cell 6 must compensate potential negative values $c_7$ and $c_8$  in cells 7 and 8. This leads to:\\
$c_1+c_2+c_3+c_5+c_6  \geq m_4+m_7+m_8$ \\
 To make a stronger procedure we remark that when $c_4 \geq 0 \land c_4+c_7 \geq 0$ then compensation for cell 8 is sufficient. We get:\\
$(c_1+c_2+c_3+c_5+c_6  \geq m_4+m_7+m_8) \lor $ \\
$ (c_4 \geq 0 \land c_4+c_7 \geq 0 \land c_1+c_2+c_3+c_5+c_6  \geq m_8 )$ \newline
The same reasoning for all other cells, where cell 10 may be omitted (we have \textit{das} $\geq$ \textit{das}).
\\
\end{proof}
We implemented this dominance test in the procedure \textit{dom()} in  ASG. 
We can also use this procedure when the two sets have different dimensions.When $e=dim(a)-dim(b)>0$ we add $e$ elements in cell 1 of pattern $b$ and then we apply the foregoing procedure. 

\begin{lemma}
$\{{\varphi{}}_1,{\varphi{}}_2,{\varphi{}}_3,{\varphi{}}_4\}$ dominates all
adequate sets, where
$$\varphi_1=(0121321110)$$
$$\varphi{}_2=(1210141110)$$
$$\varphi{}_3=(1331000301)$$
$$\varphi{}_4=(1331121110)$$
\end{lemma}
\begin{proof}
We execute ASG with $das=12$ and different values of $p,q,r$. When $p=0.7, q=0.2, r=0.1$ we find three optimal adequate sets with pattern  $\varphi_1=(0121321110).$ 
Using $p=\frac{1}{2}, q=\frac{1}{3}, r=\frac{1}{6}$, we get optimal adequate sets with pattern $\varphi{}_2=(1210141110)$.
For $p=0.35, q=0.33, r=0.32$ we obtain the optimal pattern $\varphi{}_3=(1331000301)$.
Executing ASG with  $dom(\varphi_1),dom(\varphi_2),dom(\varphi_3) $ activated (see appendix A), we get no adequate sets: $\{{\varphi{}}_1,{\varphi{}}_2,{\varphi{}}_3\}$ dominates all adequate sets when $das=12$. We get the same result when $das=13$.  
Setting $das=14$, any values of $p,q,r$ and $dom(\varphi{}_1),dom(\varphi{}_2),dom(\varphi{}_3)$ activated with $e=2$, we get 10 adequate sets (out of 50814), all dominated by or equal to $\varphi_4=(1331121110)$.
In Appendix \ref{appendix:three} we show that $\{{\varphi{}}_1,{\varphi{}}_2,{\varphi{}}_3,{\varphi{}}_4\}$ dominates all adequate sets when $das>14$.\\
{\raggedright
Conclusion:
$\{{\varphi{}}_1,{\varphi{}}_2,{\varphi{}}_3,{\varphi{}}_4\}$ dominates all
adequate sets.
}
\end{proof}
{\raggedright
Our task is now to minimize (adequate set, BAD CASES) the value of $\varphi{}$,
given values of \textit{p,q,r}.
}

Let $\Psi_i{}=1-\varphi_i$ $ (i=1,2,3,4)$. $\Psi_i$ is the probability of winning the game.

\begin{lemma}
\begin{align*}
\Psi_1&=p\left(1-2r^2\right)+{(1-p)}^2(p+r)\\
\Psi_2&=1+p^2\ r+2pr^2+p^2-p-r\\
\Psi_3&=3p(1-p-pr)\\
\Psi_4&=p(p^2-2p+2)
\end{align*}
\end{lemma}
\begin{proof}
\[
{\Psi{}}_1=1-{\varphi{}}_1=1-\left(qr^2+2q^2r+q^3+3pr^2+2pqr+pq^2+p^2r+p^2q\right).
\]
{\raggedright
Eliminate $q$ to get the desired result for $\Psi{}_1$. Analogue for $\Psi{}_2,\Psi{}_3$ and $\Psi{}_4$.
}
\end{proof}

{\raggedright
We  notice that because of $p+q+r=1$  and $p\geq{}q\geq{}r>0$, we have:\\
$p\geq{}\frac{1}{3}$ and $1-2p\leq r\leq{}\frac{1-p}{2}$ (see also Figure \ref{fig1}). }
\begin{lemma}
$\{\varphi_2,\varphi_4\}$ dominates $\varphi_1$
\end{lemma}
\begin{proof}
{\raggedright
\(
{\Psi{}}_2-{\Psi{}}_1=\left(1-p-2r\right)\left[{\left(1-p\right)}^2-2pr\right]\geq 0
\), when  $0 < r \leq min\{\frac{{(1-p)}^2}{2p},\frac{1-p}{2}\}$.\\ 
}
$\Psi_4-\Psi_1=r[2pr-(1-p)^2]\geq 0$ when $\frac{{(1-p)}^2}{2p}\leq r\leq  \frac{1-p}{2}$.\\
{\raggedright
So: ${\varphi{}}_1$ is dominated by  $\{{\varphi{}}_2,{\varphi{}}_4\}$ (see also Figure \ref{fig1}).
}
\end{proof} 
\begin{lemma}
 \({\Psi{}}_4\) is an optimal winning probability when   $$\frac{{(1-p)}^2}{2p} \leq r\leq{}\frac{1-p}{2} \ \ (\frac{1}{2} \leq p<1)$$ 
\end{lemma}
\begin{proof}
$\Psi_4-\Psi_2=(1-p-r)[2pr-(1-p)^2]=q[2pr-(1-p)^2] $
{\raggedright
$ 
{\Psi{}}_4-{\Psi{}}_3=p(p^2+p-1+3pr)> \frac{3}{2} p [2pr-(1-p)^2] $
}

and $\frac{{(1-p)}^2}{2p} \leq r\leq{}\frac{1-p}{2}$
implies $\frac{1}{2} \leq p<1$ (see also Figure \ref{fig1}).
\end{proof}
Let $\alpha$ be the solution of $\frac{1-2p}{2p}=\frac{1-p}{2}$ with $\frac{1}{3}\leq p<1$, so $\alpha=\frac {3-\sqrt{5}}{2}$.
\begin{lemma}
{\raggedright
 ${\Psi{}}_3$ is an optimal winning probability  when
}

{\raggedright
$$\left[1-2p\leq r\leq \frac{1-p}{2}\ \ \ (\frac{1}{3}\leq p \leq \alpha{}) \right ] \lor
\left[1-2p\leq r\leq \frac{1-2p}{2p}\ \ (\alpha{}\leq p \leq \frac{1}{2})\right]$$ .
}
\end{lemma}
\begin{proof}

We  have (see also Figure \ref{fig1}):\\
If  $\frac{1}{3}\leq p\leq \alpha{}$  then $r\leq \frac{1-p}{2}< \frac{1-p-p^2}{3p}$\\
If  $\alpha{}\leq p \leq\frac{1}{2}$  then  $r \leq \frac{1-2p}{2p}<\frac{1-p-p^2}{3p}$
\\
So when $\frac{1}{3}\leq p\leq  \frac{1}{2}$ we have:
\(
{\Psi{}}_3-{\Psi{}}_4=-p(p^2+p-1+3pr)>0
\).
\(
{\Psi{}}_3-{\Psi{}}_2=\left(1-2p-r\right)\left(2pr+2p-1\right)\geq 0
\) when $r\leq \frac{1-2p}{2p}$. \\
The result follows from $\frac{1-p}{2}\leq \frac{1-2p}{2p}$ when $\frac{1}{3}\leq p\leq \alpha{}$.
\end{proof}
\begin{lemma}
${\Psi{}}_2$ is an  optimal winning probability when   $$\left[\frac{1-2p}{2p} \leq r \leq \frac{1-p}{2} \ \
(\alpha{} \leq p\leq{}\frac{1}{2})\right ] \lor  \left[0 <\ r \leq \frac{{(1-p)}^2}{2p} \ \
  (\frac{1}{2}\leq{}p<1)\right]$$ 
\end{lemma}
\begin{proof}
{\raggedright
$
\Psi_2-\Psi_4=(r+p-1)[2pr-(1-p)^2]=-q[2pr-(1-p)^2]$
}

{\raggedright
\(
{\Psi{}}_2-{\Psi{}}_3=\left(2p+r-1\right)\left(2pr+2p-1\right)\).
}\\
If $p \geq \frac{1}{2}$  then $2p+r-1\leq 0$ and $ 2pr+2p-1\leq 0$ so $\Psi_2\geq \Psi_3$.

When
$p\geq \frac{1}{2}$  and $r\leq \frac{{(1-p)}^2}{2p}$ we have $\Psi_2\geq \Psi_4$.
\\
 When $\alpha\leq p\leq \frac{1}{2}$  we
have $\frac{1-2p}{2p} \leq r\leq \frac{1-p}{2}\leq \frac{(1-p)^2}{2p}$, so:
$\Psi_2\geq \Psi_4$ and $\Psi_2\geq \Psi_3$.\\ 
 See also Figure \ref{fig1}.

\end{proof}

\begin{theorem}
The next table and graphics gives a  description of optimal probabilities for asymmetric three person three color Hat Game: 

\resizebox{.9\hsize}{!}{%
\renewcommand{\arraystretch}{1.8}
     \begin{tabular}{|c|c|c|}
     \hline
Set&Region&Optimal probability\\
\hline 
  A&$\frac{{(1-p)}^2}{2p}\leq r\leq{}\frac{1-p}{2}$  $(\frac{1}{2}\leq p<1)$
         &   $\Psi_4   $    \\
  \hline
   B&$\frac{1-2p}{2p}\leq r\leq \frac{1-p}{2}$   
$\left(\alpha{}\leq p\leq \frac{1}{2}\right)\ \  \lor  \ \ 0<r\leq \frac{{(1-p)}^2}{2p} 
  \left(\frac{1}{2}\leq{}p<1\right).$
     &   $\Psi_2   $    \\
  \hline
    C&$1-2p\leq r\leq \frac{1-p}{2}\ \ \ (\frac{1}{3}\leq p\leq \alpha{}) \ \   \lor \ \ 
1-2p\leq r\leq \frac{1-2p}{2p}\ \ (\alpha{}\leq{}p\leq \frac{1}{2})$
     &        $  \Psi_3 $ \\
  \hline
     \end{tabular}% 
}
\begin{align*}
\alpha&=\frac{3-\sqrt{5}}{2}\\
\Psi_2&=1+p^2r+2pr^2+p^2-p-r\\
\Psi_3&=3p(1-p-pr)\\
\Psi_4&=p(p^2-2p+2)
\end{align*}

\begin{figure}
\begin{center}
\includegraphics[scale=0.7]{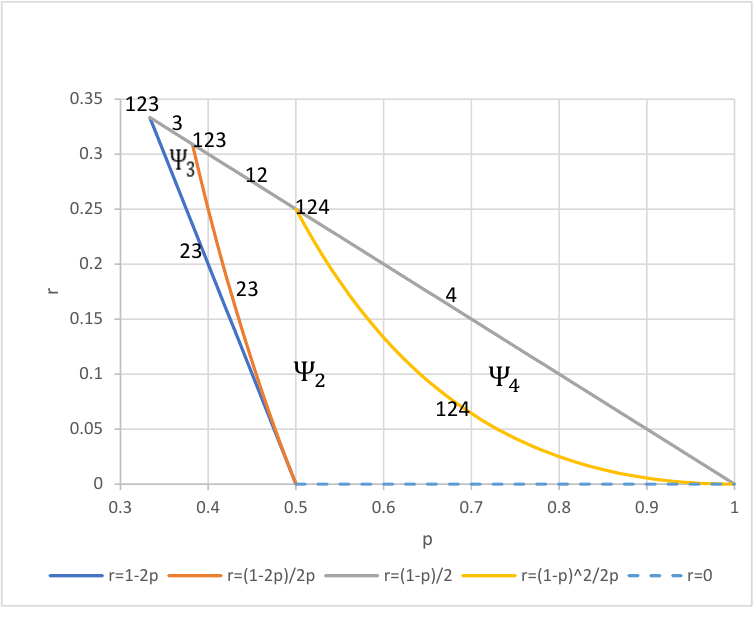}
\end{center}
\caption{Optimal probabilities}
\label{fig1}
\end{figure}
\begin{center}
\includegraphics[scale=0.7]{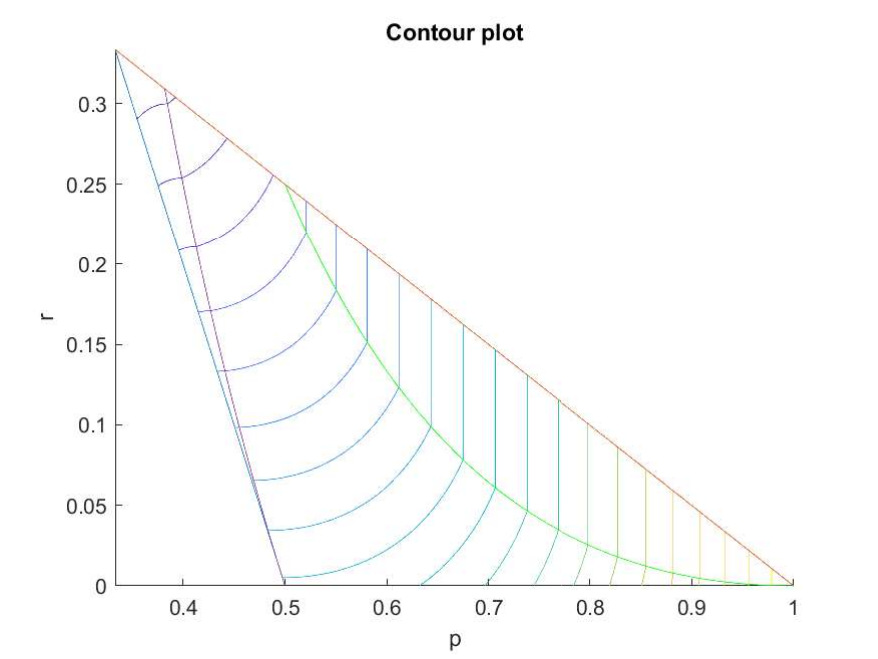}
\end{center}
\begin{center}
\includegraphics[scale=0.6]{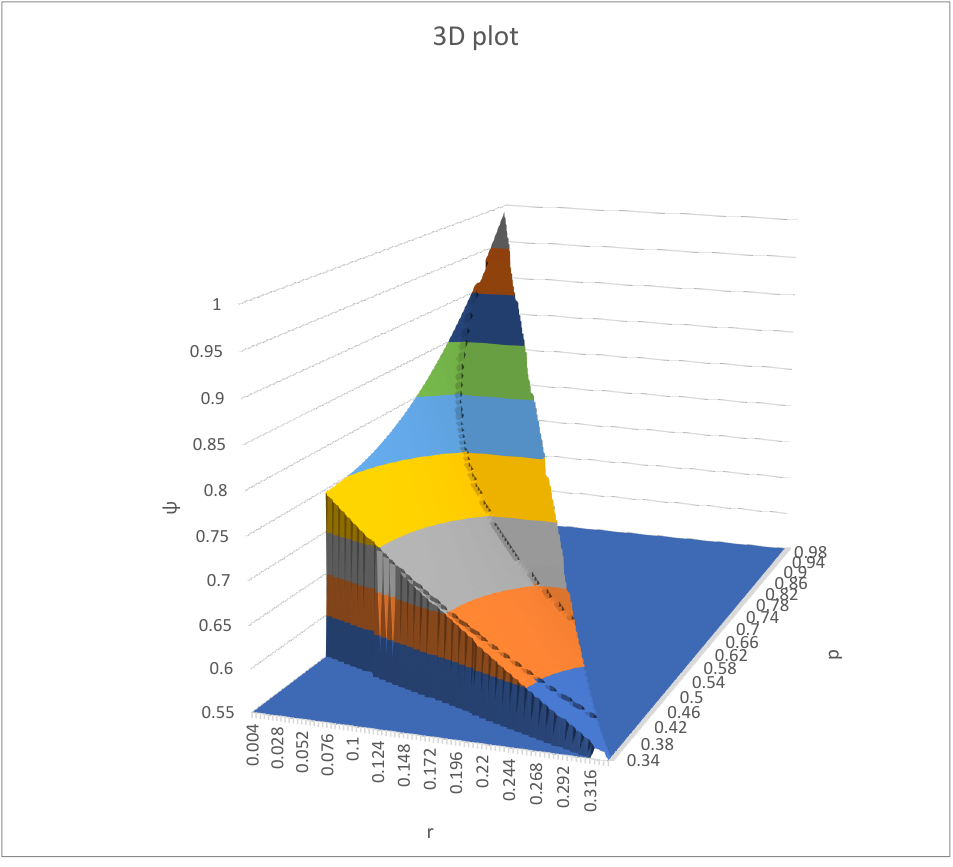}
\end{center}
\end{theorem}
\begin{proof}
Use Lemma's 6-9.
\end{proof}

We find all solutions of the asymmetric three person \textit{two} color Hat Game by taking $r=0$:
\begin{theorem}
If $p>\frac{1}{2}$ then we have optimal probability $1+p^2-p$ with optimal decision matrix: 
\begin{center}
    \begin{tabular}{|cccc|}
    \hline
   {00} & {01} &  10    & 11      \\
   \hline
   
    0     &       &             & 1           \\
          & 0          & 1     &       \\
          & 0         & 1     &   \\
    
          \hline
          
    \end{tabular}%
    \end{center}
If $p=q=\frac{1}{2}$ then we have optimal probability $\frac{3}{4}$ and two optimal decision matrices: the one already found when $p>\frac{1}{2}$ and the well-known:
\begin{center}
    \begin{tabular}{|cccc|}
    \hline
   {00} & {01} &  10    & 11      \\
   \hline
   
    1     &       &             & 0           \\
       1   &           &      &    0   \\
       1  &          &      &  0 \\
    
          \hline
          
    \end{tabular}%
    \end{center}
\end{theorem}
\begin{proof}
If $p>\frac{1}{2}$ then we are in $B$ with optimal probability $\Psi_2=1+p^2-p$ and we get the optimal decision matrix using the result of region $B$ in section 3.2 and deleting color 2. \\
If $p=q=\frac{1}{2}$ then we have to do with sets $B$ and $C$. We get $\Psi_2=\Psi_3=\frac{3}{4}$ and a second optimal decision matrix by deleting color 2 in the decision matrix of region C in section 3.2.   
\end{proof}
The same result is obtained in ~\cite{TU}.\\
We investigate the behavior of the optimal probability $\Psi_{opt}$ when $p \rightarrow 1$:
\begin{theorem}
$\Psi_{opt}=1-(1-p)+(1-p)^2+\mathcal{O}([1-p]^3) \quad (p \rightarrow 1) $
\end{theorem}
\begin{proof}
When $\frac{(1-p)^2}{2p}\leq r \leq \frac{1-p}{2}$ we have: $\Psi_{opt}=\Psi_4=p(p^2-2p+2)=1-(1-p)+(1-p)^2-(1-p)^3$.\\
When $0\leq r \leq \frac{(1-p)^2}{2p} $ then $\Psi_{opt}=1-p+p^2+p^2r+2pr^2-r$, where 
$\lvert p^2r+2pr^2-r \rvert \leq 2pr^2+(1-p^2)r \leq \frac{(1-p)^4}{2p}+\frac{(1-p^2)(1-p)^2}{2p}=\frac{(1-p)^3}{p}=$ \\
$\frac{(1-p)^3}{1+(p-1)}=(1-p)^3+\mathcal{O}([1-p]^4)\quad (p \rightarrow 1)$.
\end{proof}
\subsection{Optimal strategies}
\begin{definition}
Two adequate sets are  \textit{isomorphic} when there is a permutation of the three players which projects one set to another.
\end{definition}
\begin{theorem}
Optimal strategies for three person, three color, hat game are:\\
Region A:\\
\begin{center}
    \begin{tabular}{|ccccccccc|}
    \hline
   {00} & {01} & {02} & 10    & 11    & 12    & 20    & 21    & 22 \\
   \hline
    0     &       &       &       &   $\star$     & $\star$       &       &      $\star$  &  $\star$ \\
          & 0     & 0     &       & 0     & 0     &       & 0     & 0 \\
          & 0     & 0     &       & 0     & 0     &       & 0     & 0 \\
          \hline        
    \end{tabular}%
 \end{center}
 Region B:\\
 \begin{center}
    \begin{tabular}{|ccccccccc|}
    \hline
   {00} & {01} & {02} & 10    & 11    & 12    & 20    & 21    & 22 \\
   \hline  
    0     &       &       &       & 1     & 1     &       & 1     & 1 \\
          & 0     & 0     & 1     &       &       &       & 0     & 0 \\
          & 0     & 0     & 1     &       &       &       & 0     & 0 \\    
          \hline          
    \end{tabular}%
    \end{center}
Region C:\\
    \begin{center}
    \begin{tabular}{|ccccccccc|}
    \hline
   {00} & {01} & {02} & 10    & 11    & 12    & 20    & 21    & 22 \\
   \hline
    1     &       &       &       & 0     & 0     &       & 0     & 0 \\
    1     &       &       &       & 0     & 0     &       & 0     & 0 \\
    1     &       &       &       & 0     & 0     &       & 0     & 0 \\      
          \hline          
    \end{tabular}%
\end {center}  
\end{theorem}
\begin{proof}
{\raggedright
We start with region A.
}
{\raggedright
We take a point in A (e.g.  $\textit{p}=0.7,\ q=0.2,\ \ r=0.1 $  ). ${\Psi{}}_4$ is winner in A, so $\textit{das}$=14. Execute the program ASG with $\textit{das}$=14    and we find
three optimal adequate sets:
}
\begin{center}
    \begin{tabular}{rrrrrrrrrrrrrr}
    4     & 5     & 7     & 8     & 9     & 13    & 14    & 16    & 17    & 18    & 22    & 23    & 25    & 26 \\
    1     & 2     & 12    & 13    & 14    & 15    & 16    & 17    & 21    & 22    & 23    & 24    & 25    & 26 \\
    3     & 6     & 10    & 11    & 13    & 14    & 16    & 17    & 19    & 20    & 22    & 23    & 25    & 26 \\
    \end{tabular}% 
\end{center}
These sets are isomorphic: cycles $(13)$ and $(23)$ will project the first set to the second and third one (using ternary code).
{\raggedright
Program DMG with $\textit{das}$=14  gives 
the  optimal decision matrix belonging to the first adequate set in region A.
}

    {\raggedright
Region B:
}
{\raggedright
We consider a point in B (e.g.  $\textit{p}=\frac{1}{2},\ q=\frac{1}{3},\ r=\frac{1}{6}$ 
) , execute the program ASG    with \textit{das}=12 (${\Psi{}}_2\ $is
winner) and we get 3 optimal adequate sets:
}
\begin{center}
    \begin{tabular}{rrrrrrrrrrrrrr}
   
    4     & 5     & 7     & 8     & 9     & 11    & 15    & 18    & 22    & 23    & 25    & 26 \\
    1     & 2     & 7     & 12    & 14    & 15    & 17    & 19    & 21    & 23    & 24    & 26 \\
    3     & 5     & 6     & 10    & 11    & 16    & 17    & 19    & 20    & 21    & 25    & 26 \\
    \end{tabular}%
 
\end{center}
The three sets are isomorphic (again $(13)$ and $(12)$ will do).
{\raggedright
DMG  gives 
the  optimal decision matrix belonging to the first adquate set of region B.
}

{\raggedright
Region C:
}

{\raggedright
\textit{das}=12 (${\Psi{}}_3\ $is winner). Using e.g.   $\textit{p}=0.35,\ q=0.33\ ,\
r=0.32$  we get one optimal adequate set when executing the program ASG with \textit{das}=12: 
}
\begin{center}
    \begin{tabular}{rrrrrrrrrrrr}
    0     & 2     & 6     & 13    & 14    & 16    & 17    & 18    & 22    & 23    & 25    & 26 \\
    \end{tabular}%

    \end{center}
    
{\raggedright
Procedure DMG   gives 
the  optimal decision matrix for region C.
}

\end{proof}
\subsection{Computational complexity}
{\raggedright
We consider the number of strategies to be examined to solve the hat problem
with $N$ players and q colors. Each of the $N$players has $q^{N-1}$ possible
situations to observe and in each situation there are q+1  possible guesses. So
we have ${({(q+1)}^{q^{N-1}})}^N$ possible strategies. Krzywkowski [14] shows
that is suffices to examine ${({(q+1)}^{q^{N-1}-1})}^N$ strategies.
}

{\raggedright
The adequate set method has to deal with \{$i_1,i_2,..,i_{\textit{das}}\}$ with
$0\leq{}i_1<i_2<..<i_{\textit{das}}\leq{}q^N-1$.
}

{\raggedright
The number of strategies for fixed \textit{das} is the number of subsets of dimension \textit{das} of
\{0,1,\ldots{},$\ q^N$ -1\}: $\left(\begin{array}{l}q^N \\
\textit{das}\end{array}\right)$.
}
But we have to test all possible values of \emph{\textit{das}}. So the correct expression is: $\sum_{\emph{\textit{das}}}
\left(\begin{array}{l}q^N \\
\textit{das}\end{array}\right)=2^{(q^N)}$.
{\raggedright
The power of the adequate set method in the asymmetric 3 person, 3 color game is
shown in the next table of computational complexity:
}

\begin{center}
     \begin{tabular}{|ccc|}
     \hline
brute force&	Krzywkowski 	&adequate set method\\
1,80144E+16&	2,81475E+14&	134217728\\
  \hline
    \end{tabular}% 
\end{center}

\section{Symmetric Hat Game}
In this section we focus on the symmetric Hat Game with three players and three colors. 
\subsection{Symmetric three color three person Hat Game}
\begin{definition}
Two adequate sets are \textit{equivalent} when they induce the same probability function.
\end{definition}
\begin{definition}
The \textit{order} of an adequate set is the number of equivalent adequate sets.
\end{definition}
\begin{theorem}
When $\textit{p}=q=r=\frac{1}{3}$ the optimal probability is $\frac{5}{9}$ and we have a collection of 77 non-isomorphic optimal decision matrices, each of order 1,3 or 6. We give one matrix of each order:\\
Order 1, adequate set \{0	1	3	4	8	9	10	12	13	20	24	26\}

\begin{center}
    \begin{tabular}{|ccccccccc|}
    \hline
   {00} & {01} & {02} & 10    & 11    & 12    & 20    & 21    & 22 \\
   \hline
     2    &   2    &       &      2 &    2   &       &       &       &  1\\
    2    &   2    &       &      2 &    2   &       &       &       &  1\\
    2    &   2    &       &      2 &    2   &       &       &       &  1\\
          \hline
          
    \end{tabular}%
 \end{center}
 Order 3, adequate set \{4	5	7	8	9	13	14	16	17	18	20	24\}

\begin{center}
    \begin{tabular}{|ccccccccc|}
    \hline
   {00} & {01} & {02} & 10    & 11    & 12    & 20    & 21    & 22 \\
   \hline
    0     &       &       &       &  2     & 2      &       &      2 & 2 \\
          &  0    &    0  &       &   0   & 0     &   1    &      &  \\
             &  0    &    0  &       &   0   & 0     &   1    &      &  \\
          \hline
          
    \end{tabular}%
 \end{center}
 Order 6, adequate set \{4	5	7	8	9	13	14	16	17	18	19	24\}

\begin{center}
    \begin{tabular}{|ccccccccc|}
    \hline
   {00} & {01} & {02} & 10    & 11    & 12    & 20    & 21    & 22 \\
   \hline
     0	  &       &       &       &    2   &  2     &       &      2 & 2 \\
          &    0  &  0    &       &    0  &  0    &  1     &      &  \\
          &    0  &  0    &       &    0  &  0    &  2     &      &  \\
          \hline
          
    \end{tabular}%
 \end{center}
 \end{theorem}
\begin{proof}
Running ASG  with any values of $p,q,r$ gives 324 adequate sets. When $\textit{p}=q=r=\frac{1}{3}$ all these sets are optimal with probability $\frac{5}{9}$. 
 We run ASG with $p=0.71, q=0.23, r=0.06$ (we don't want contamination of the probabilities). We sort the adequate sets by probability and find 75 equivalent sets of order 1,3,6 or 12 (see Appendix \ref{appendix:five1}). 
 \begin{center}
        \begin{tabular}{|c|c|c|}
     \hline
order&count&order $\star$ count\\
\hline 
  1& 6
         &   6   \\
  \hline
   3&36 
     &  108    \\
  \hline
    6&31

     & 186 \\
  \hline
    12&2
         &  24    \\
  \hline
  \hline
 sum  &75
     &      324 \\
     \hline
           \end{tabular}%
 \end{center}
 Each set of order 12 can be split  in two sets of order 6 which are not isomorphic:  count number of non passes for each player, which gives a 663 or 555 pattern. Each set of order 3 is isomorphic to 2 equivalent sets of order 3 and each set of order 6 is isomorphic to 5 equivalent sets of order 6, which can be verified by permutations of the players. 
In total we have 77 non-isomorphic adequate sets.
All 324 decision matrices are generated by DMG.  
\end{proof}
\appendix
\clearpage

\includepdf[
    pages=1,
    scale=0.7,
    nup=1x1,
    frame,
    offset={2.5cm, -1.0cm},
    pagecommand={%
        \section{Adequate Set Generator}
	\label{appendix:one}
    %\appendix
}       
]{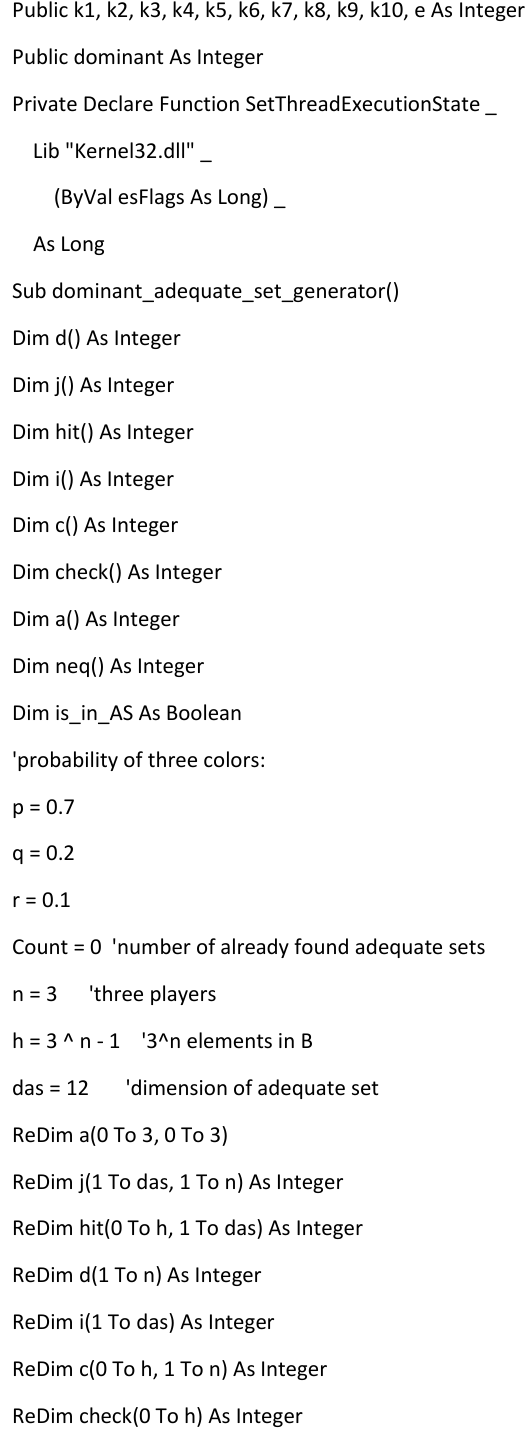}

\includepdf[
    pages=2-5,
    scale=0.7,
    nup=1x1,
    frame,
    offset={2.5cm, -1.0cm},
    pagecommand={%
       %\appendix
}       
]{ASGe.pdf}

\includepdf[
    pages=1,
    scale=0.7,
    nup=1x1,
    frame,
    offset={2.5cm, -1.0cm},
    pagecommand={%
        \section{Decision Matrix 		generator}
	\label{appendix:two}
    %\appendix
}       
]{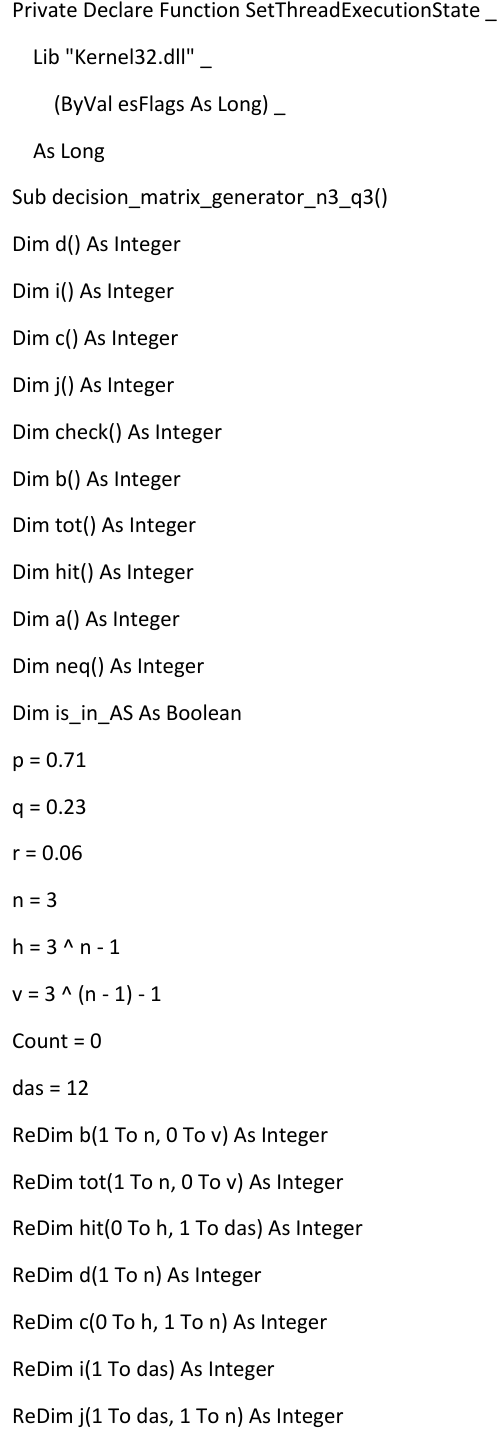}

\includepdf[
    pages=2-3,
    scale=0.7,
    nup=1x1,
    frame,
    offset={2.5cm, -1.0cm},
    pagecommand={%
       %\appendix
}       
]{DMG.pdf}

\section{Dominance when das > 14}
\label{appendix:three} 
Running  ASG with $das=15$ and procedure $dom()$ activated for $\varphi_1,\varphi_2,\varphi_3,\varphi_4$ with $e=3$, we get 70 sets. Sorting by probability gives 12 different probability classes, all dominated by or equal to 4 sets:
\begin{center}
  \begin{tabular}{|cccccccccc|}
  \hline
 0&	1&	2&	1&	3&	5&	2&	0&	0&	1\\
1&	1&	1&	1&	2&	6&	2&	0&	0&	1\\
0&	2&	2&	0&	3&	4&	3&	0&	0&	1\\
1&	2&	1&	0&	2&	5&	3&	0&	0&	1\\
    \hline
    
    \end{tabular}%
\end{center}
When $das=16$ and activating $dom()$ for $\varphi_1,\varphi_2,\varphi_3,\varphi_4$ with $e=4$ and the 4 sets just found in $das=15$,  we get 4 sets:
\begin{center}
  \begin{tabular}{|cccccccccc|}
  \hline
 0	&2&	3	&1&	3&	4&	2&	0&	0&	1\\
1&	2&	2&	1&	2&	5&	2&	0&	0&	1\\
0&	3&	3&	0&	3&	3&	3&	0&	0&	1\\
1&	3&	2&	0&	2&	4&	3&	0&	0&	1 \\
    \hline
    
    \end{tabular}%
\end{center}
Executing ASG with $das=17$ and activating $dom()$ for $\varphi_1,\varphi_2,\varphi_3,\varphi_4$, the 4 sets  found in $das=15$ and the 4 sets in $das=16$, we get 9 sets all with pattern:
1	3	3	1	2	4	2	0	0	1
\newline
When $das \geq 18$ and activating $dom()$ for $\varphi_1,\varphi_2,\varphi_3,\varphi_4$, the 4 sets  found in $das=15$, the 4 sets in $das=16$ and the pattern of $das=17$ we get only patterns which are dominated.
\newline
The 9 patterns we found in $das=15, 16, 17$ are dominated by $\varphi_2$ (except the first one which is dominated by $\varphi_1$);
 use domination in:
\begin{center}
  \begin{tabular}{|cccccccccc|}
  \hline
    00    & 01    & 02    & 03    & 10    & 11    & 12    & 20    & 21    & 30 \\
            &     &    &     &  &    & 2    & 0    & 0   & 1 \\
        &     &    &     &  &    & 1    & 1    & 1    & 0 \\
    \hline
    
    \end{tabular}%
\end{center}
We have: $pq^2-p^2r-p^2q+p^3=p[(p-q)^2+p(q-r)] \geq 0$.
\clearpage

 \section{75 probability classes; p=0.71, q=0.23, r=0.06 }
\label{appendix:five1} 
 
\includegraphics[width=\textwidth]{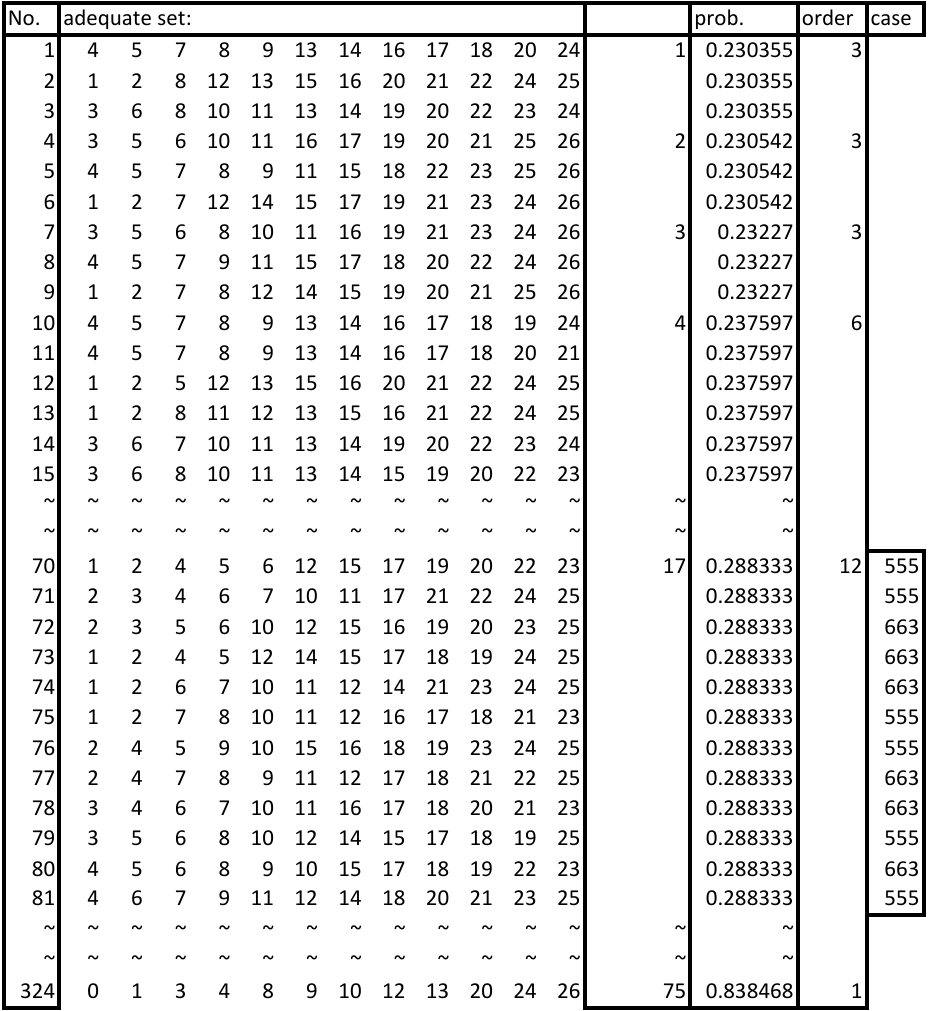}

%BIBLIOGRAPHY
% You do not have to use the same format for your references, but 
%    include everything in this file.  Don't use natbib please.
% If you use BibTeX to create a bibliography, copy the .bbl file into here.
% \newblock is optional (it adds a little space)


\clearpage
\begin{thebibliography}{99}


\bibitem{WB}	W. Blum, Denksport f\"{u}r Huttr\"{a}ger, Die Zeit, May 3, 2001. 

\bibitem{JB}	J. Buhler, Hat tricks, Math. Intelligencer 24 (2002), 44-49.

\bibitem{EB}	E. Burke, S. Gustafson, G. Kendall, A puzzle to challenge genetic programming, Genetic Programming, 136-147, Lecture Notes in Computer Science, Springer, 2002.

\bibitem{SB}	S. Butler, M. Hajianghayi, R Kleinberg, T. Leighton, Hat guessing games, SIAM J Discrete Math 22 (2008), 592-605.

\bibitem{GC}	G. Cohen, I. Honkala, S. Litsyn, A. Lobstein, Covering Codes, North-Holland, Mathematical Library 54, 1997.

\bibitem{TE}	T. Ebert, Applications of recursive operators to randomness and complexity, Ph.D. Thesis, University of California, Santa Barbara, 1998.

\bibitem{EMV}	T. Ebert, W. Merkle, H. Vollmer, On the autoreducibility of random sequences, SIAM J. Comp. 32 (2003), 1542-1569.

\bibitem{MG}	M. Gardner, The 2nd Scientific American Book of Mathematical Puzzles \& Diversions, Simon and Schuster, New York, 1961.

\bibitem{GU} Guo, W., Kasala, S., Rao, M.B., Tucker, B. (2006). The Hat Problem and Some Variations. In: Balakrishnan, N., Sarabia, J.M., Castillo, E. (eds) Advances in Distribution Theory, Order Statistics, and Inference. Statistics for Industry and Technology. Birkhäuser Boston.

\bibitem{CH}	C. Hardin, A. Taylor, The Mathematics of coordinated inference: A study of generalized hat problems, Springer International Publishing Switzerland 2013.

\bibitem{BJ}	B. Johnson, \href {http://mathstat.slu.edu/~johnson/public/maths/hatproblem.pdf}{http://mathstat.slu.edu/~johnson/public/maths/hatproblem.pdf}

\bibitem{GK}	G. Kéri, Tables for bounds on covering codes, www.sztaki.hu/~keri/codes
                                                                                                                                                                                                                                                                                                                                                                                                                   
\bibitem{MK}	M. Krzywkowski, On the hat problem, its variations, and their applications,
Annales Universitatis Paedagogicae Cracoviensis Studia Mathematica 9 (1), 55-67, 2010.

\bibitem{MKR}	M Krzywkowski, A more colorful hat problem, Annales Universitatis Paedagogicae Cracoviensis Studia Mathematica 10 (1), 67-77, 2011.

\bibitem{HL}	H. Lenstra, G. Seroussi, On hats and other covers, (Extended Summary),  Arxiv {cs/0509045v1} [cs.IT] 15 sep 2005.

\bibitem{JP}	J. Poulos, Could you solve this 1 million hat trick?, abcNews, November 29, 2001

\bibitem{SR} 	S. Robinson, Why mathematicians now care about their hat color, The New York Times, Science Times Edition,	page D5, April 10, 2001.

\bibitem{TU} T. van Uem, A Generalized Hat Game, Arxiv 170404244.v6 [math.CO] 16 mar 2023.

\bibitem{TA}  Uthaipon Tantipongpipat. A combinatorial approach to Ebert’s hat game with many colors. Electron. J. Combin., 21(4):Paper 4.33, 18, 2014.

\bibitem{PW}	P. Winkler, Mathematical Mind-Benders, A.K. Peters, Wellesley, Massachusetts, 2007


\end{thebibliography}
\end{document}